# Measurement of the vertical non-uniformity of the plasma sheath in a complex plasma

Jie Kong, Truell W. Hyde, Brandon Harris, Ke Qiao, and Jorge Carmona-Reyes

*Abstract*— Employing an attenuated oscillation method, the anisotropic interaction force between two vertically aligned dust particles located in the sheath of a complex plasma was measured experimentally based on a linear approximation to the interaction force. Experimental data shows that although both particles experience a repulsive interaction force, the upper particle experiences a stronger magnitude force than does the lower. This result can be explained by the ion wakefield since the lower particle resides within the ion wakefield generated by the streaming ions passing through the sheath and around the upper particle.

*Index Terms*— Complex Plasma, Ion Wakefield, Diagnostics, Plasma Sheath, Anisotropic Effects.

## I. INTRODUCTION

DUST particles within a complex plasma sheath are acted upon by a variety of forces such as the electrostatic, gravitational, and ion drag forces. Under appropriate conditions, vertical chains of dust particles are often seen to form in laboratory dusty plasma experiments with the formation of two- or three-particle chains (or longer) possible. In this case, the ion wakefield force plays an important role in establishing overall equilibrium within the complex plasma.

When a vertical dust chain forms, the interaction between particles is anisotropic, primarily due to the ion focusing effect, or wakefield. Both theory and numerical simulation have shown that the particle downstream to the ion flow is under the influence of the wakefield from the upper particle [1] – [6]. The charge on each of the two particles is also in general slightly different and dependent on their separation distance [7], [8]. Experimentally Takahashi, et al [9] proved that under horizontal perturbation, the upper particle's motion can influence the motion of the lower particle but not vice versa. Takahashi proved this by employing a laser to horizontally perturb the upper particle in a vertical dust chain and then showing that its subsequent motion created a measurable effect on the lower particle; perturbing the lower particle in an identical manner resulted in no such effect on the upper particle. More recently, Hebner, et al [10], [11] measured the interaction forces directed along a vector connecting two particles of differing radii located within a potential well. By examining the trajectory of each particle as they approached one another, it was determined there were measurable attractive forces in the horizontal direction.

However to date, no such examination of the interaction forces in the vertical direction has been conducted. In order to properly examine the anisotropic properties of the plasma sheath, a direct experimental measurement of the interaction force along the vertical direction between the dust particles comprising a vertical chain is essential.

The simplest system for studying the aforementioned anisotropic behavior consists of a two-particle vertical chain. Experimentally, such two-particle vertical pairs are a commonly seen configuration within laboratory dusty plasmas while theoretically they are relatively simple to analyze, even exhibiting exact solutions in some cases.

Additionally, vertical oscillations generated in such two-particle systems alter the relative distance between the particles, generating a variation in the interparticle force between them. The resulting perturbation of this interaction force will in turn influence overall system behavior. Therefore by studying individual vertical oscillations of the particles within the chain, information about both the interparticle interaction force and the anisotropic properties of the plasma sheath can be derived.

In the following, an experimental method based on a linear approximation to the interaction force is introduced. This method is then employed to produce and then examine attenuated oscillations in the vertical direction of a two-particle chain. Using a fast Fourier transformation technique, the resulting spectrum is analyzed and information concerning the individual interaction forces obtained.

## II. Description of the theory

For a vertical oscillating pair of dust particles, the general equations of motion can be expressed as,

$$\ddot{y}_1 + \beta_1 \dot{y}_1 + \omega_{01}^2 y_1 = f_{01} \sin \omega t + f_{21}$$
$$\ddot{y}_2 + \beta_2 \dot{y}_2 + \omega_{02}^2 y_2 = f_{02} \sin \omega t + f_{12} \quad (1)$$

In the above, $y_{1,2}$ represent the vertical positions of particles 1 and 2 respectively (where 1 represents the upper and 2 the lower particle), $\beta_{1,2}$ are their neutral drag coefficients, $\omega_{01,2}$ their resonant frequencies, $\omega$ is the applied external driving frequency, $f_{12} = F_{12}/m_2$, $f_{21} = F_{21}/m_1$, $m_{1,2}$ are the dust particle masses, and $F_{12,21}$ are the interaction forces between the two particles, which is a function of their separation distance $|y_1 - y_2|$. For small oscillation amplitudes, the interaction force can be assumed linear to $|y_1 - y_2|$, i.e. a quadratic potential. Making the further assumption that external driving forces are negligible, equation (1) can now be written as

$$\ddot{y}_1 + \beta_1 \dot{y}_1 + \omega_{01}^2 y_1 = \gamma_1 (y_2 - y_1)$$
$$\ddot{y}_2 + \beta_2 \dot{y}_2 + \omega_{02}^2 y_2 = \gamma_2 (y_1 - y_2) \quad (2)$$

where $\gamma_{1,2}$ are constants which can be either positive or negative; in this case they are positive for a repulsive force and negative for an attractive force. Fourier transformation of equation (2) yields,

$$\begin{cases}(-\omega^2 + i\omega\beta_1 + \omega_{01}^2) Y_1 = \gamma_1 (Y_2 - Y_1) \\ (-\omega^2 + i\omega\beta_2 + \omega_{02}^2) Y_2 = \gamma_2 (Y_1 - Y_2)\end{cases} \quad (3)$$

To solve equation (3), we first make the assumption that $Y_2 = R_1 e^{i\delta_1} Y_1$, and for the purpose of symmetry $Y_1 = R_2 e^{i\delta_2} Y_2$, where $R_{1,2} = R_{1,2}(\omega)$ and $\delta_{1,2} = \delta_{1,2}(\omega)$. (Both can be determined experimentally.) Under this assumption, Eq. (3) becomes,

$$\frac{1}{\gamma_1 R_1} \sqrt{(\omega_{01}^2 - \omega^2 + \gamma_1)^2 + \omega^2 \beta_1^2} \cdot e^{i\Delta_1} = e^{i\delta_1}$$
$$\frac{1}{\gamma_2 R_2} \sqrt{(\omega_{02}^2 - \omega^2 + \gamma_2)^2 + \omega^2 \beta_2^2} \cdot e^{i\Delta_2} = e^{i\delta_2} \quad (4)$$
$$\tan \Delta_{1,2} = \frac{\omega \beta_{1,2}}{\omega_{01,2}^2 - \omega^2 + \gamma_{1,2}}$$

Solving for the force constants $\gamma_{1,2}$ in equation (4) yields,

$$\gamma_{1,2} = -\frac{\omega_{01,2}^2 - \omega^2}{1 - R_{1,2}^2}$$
$$\pm \frac{\sqrt{(\omega_{01,2}^2 - \omega^2)^2 - (1 - R_{1,2}^2)((\omega_{01,2}^2 - \omega^2)^2 + \omega^2 \beta_{1,2}^2)}}{1 - R_{1,2}^2} \quad (5)$$

In the above, both $\omega_{01,2}$ and $\beta_{1,2}$ can be obtained experimentally. $R_{1,2}$ are defined as the ratio of the oscillation amplitudes in $\omega$-space, and are experimentally measurable as well. For an isotropic plasma, $F_{21} = f_{21} m_1$ which should equal $F_{12} = f_{12} m_2$. Therefore, measuring $\gamma_{1,2}$ allows the anisotropic properties of the ion focusing effect in the vertical direction to be derived.

The critical assumption allowing derivation of equation (5) is the linear interaction force approximation, $f_{1,2} = \gamma_{1,2} |y_2 - y_1|$, which requires small amplitude oscillations. In order to minimize the nonlinear effects, it is therefore important to maintain low neutral gas pressure and control the oscillation amplitude.

## III. Experimental Procedure and results

To validate the theory given above, an experiment was conducted using the CASPER GEC rf reference cell at Baylor University [12]. In the CASPER GEC cell, a radio-frequency, capacitively coupled discharge is formed between two parallel-plate electrodes, 8 cm in diameter and separated by 3 cm, with the bottom electrode air-cooled. The lower electrode is powered by a radio-frequency signal generator, while the upper electrode is grounded as is the chamber. To provide confinement of the dust particles, the bottom electrode has a circular cutout of 2.5 cm diameter. The signal generator is coupled to the electrode through an impedance matching network and a variable capacitor attenuator network. Measured positions for melamine formaldehyde particles of diameter 8.89 ± 0.09 μm were recorded using a CCD camera at 120 frames per second for an argon gas plasma held at 100 mTorr under a rf power of 5W and frequency 13.56 MHz. The CASPER plasma discharge apparatus is described in greater detail in [13]. Representative dusty plasma parameters as measured experimentally during this experiment are given in Table 1.

Under the described conditions, dust particles formed a single primary layer exhibiting hexagonal crystal structure along with several vertical particle pairs immediately below this layer. These vertical pairs were illuminated employing a vertically fanned laser beam, and their motions monitored and recorded employing a side-mount camera with a capture rate of 120 fps.

TABLE I
Measured Plasma Parameters

| | | |
|---|---|---|
| $n_e$ [a] | electron density | $2.5 \times 10^8$ (cm$^{-3}$) |
| $T_e$ [a] | electron temperature | 3.3 eV |
| $\lambda_D$ [†] | Debye length | 100 μm |
| $V_n$ | natural DC bias | -29 V |
| $\Delta V$ | step DC voltage | -11 V, -5V |
| $P_{RF}$ | driving RF power | 5W |
| $f_{RF}$ | driving RF frequency | 13.56 MHz |

[a] Langmuir probe measurement result.
[†] Debye length derived using the technique shown in [16].

In order to generate vertical oscillations without using an external periodic driving force, a two-step process was employed. First, the dust particles were raised a distance (ΔH) above their natural equilibrium positions through the addition of an external DC bias provided to the lower electrode. (For example, inserting an external DC bias of −5V from the natural bias raises the particles a distance ΔH = 150μm above their natural equilibrium position.) The particles were then allowed to return to their natural equilibrium positions by removing the external bias. This process created a series of attenuated oscillations around the particles' natural equilibrium positions. Oscillations created in this manner resulted in small oscillation amplitudes allowing nonlinear effects to be ignored. A Fourier transform was performed on the resulting oscillation data in order to obtain frequency distribution spectrum. This procedure was then repeated for various values of ΔH both above and below the particles' natural equilibrium positions.

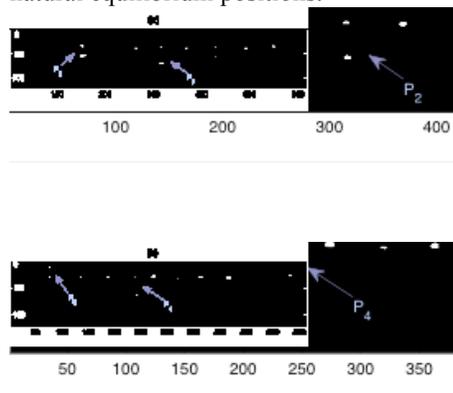

Fig. 1. Measured dust particle vertical positions. (a) Lower particles in both p1 and p2 are below the primary layer while (b) the top particle in particle-pair p3 is above the primary layer, and the lower particle in pair p4 is below the primary layer. The vertical separation distances for each pair are 21, 35, 20, 35 pixels for p1 through p4 respectively. The axes for both X and Y are in pixels with the measured pixel size approximately 12 μm.

In Fig. 1, experimentally measured vertical particle positions can be seen. As shown, both pairs in (a) have lower particles beneath the primary layer, while in (b) one pair (p3) has a particle above while the other pair (p4) has one below the primary layer. Separation distances between the two particles are different for each pair.

Fig. 2 shows the raw oscillation data along with the resultant Fourier transformation spectra. (In all cases, the frequency step size for the FFT analysis shown is 0.1 Hz while the estimated frequency resolution (Δf) is approximately 0.5 The measured value of the background noise level in the FFT spectrum for both the upper and bottom particles is approximately 0.01.) As shown, once released particles returned to their original equilibrium position within one second. Oscillation amplitudes for the upper and lower particles are attenuated at different rates; also, the oscillation amplitudes of the upper particles decreased more rapidly than did that of the lower particles. Particle pairs with larger particle separation distances, for example such as those found in p2 and p4, exhibit a single peak in their upper particle FFT spectra; particle pairs having smaller separation distances such as those seen in p1 and p3, produce an upper particle spectra exhibiting two distinct peaks.

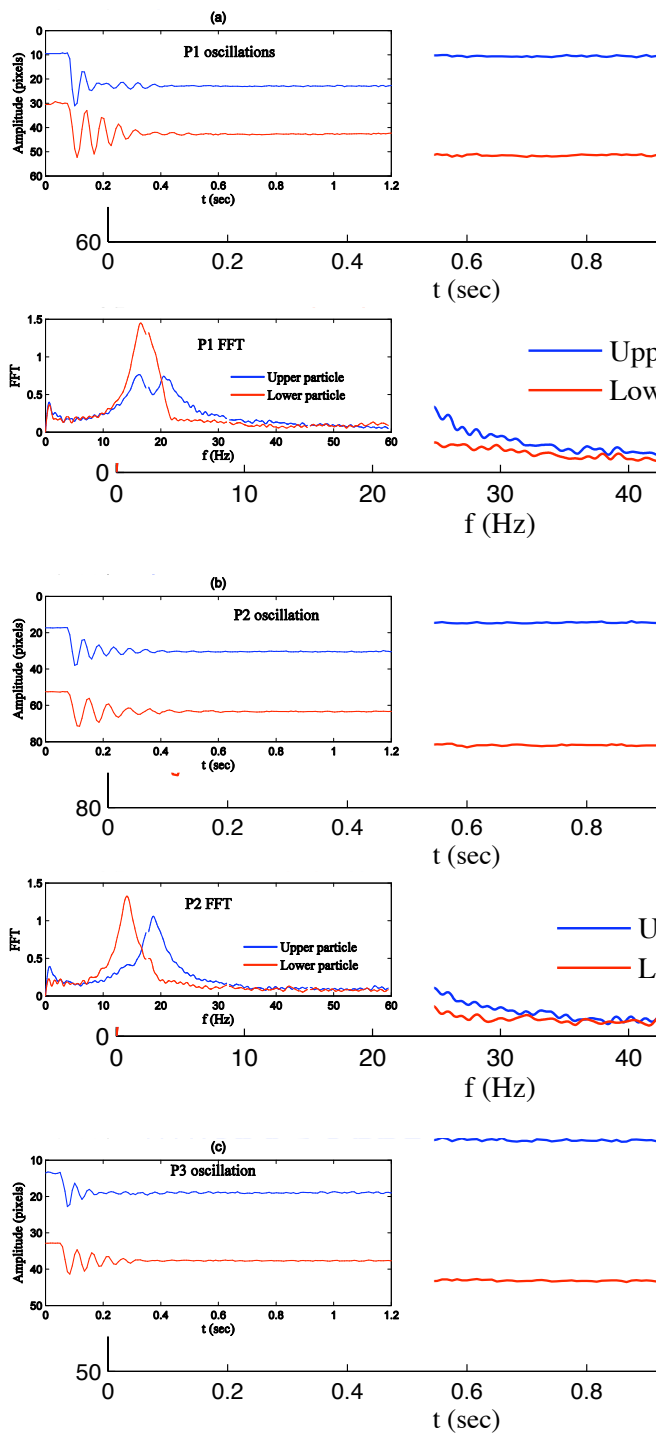

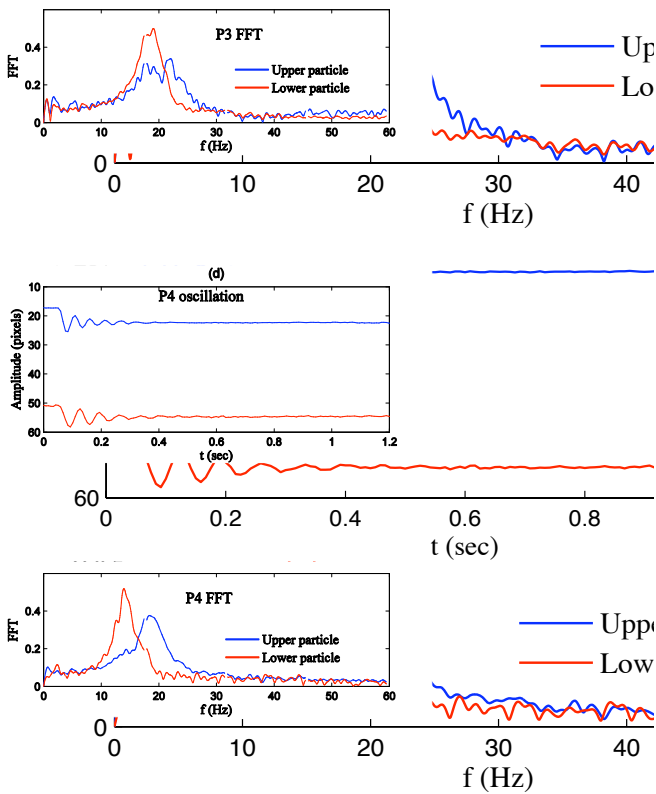

Fig. 2. Figure 2(a) through 2(d) show the oscillation amplitudes and resulting FFT spectra for p1 – p4. For larger separations, p2 and p4, the FFT spectra show a single peak for the upper particles, while the smaller separations, p1 and p3, their spectra show two distinct peaks for the upper particles.

As shown in Fig. 1, the upper particles in p1 and p2 and the lower particle in p3 and the upper particle in p4 are at the same height above the lower electrode. According to Equation 1, particles at the same height should have the same (or very close to the same) resonance frequency. The experimental data represented in Figure 2 are in agreement with this prediction as illustrated by the fact that the upper particles in p1 and p2 as well as the lower particle in p3 and the upper particle in p4 all have peaks around 19 Hz. Since the vertical interaction force between particles is a function of their separation distance, the larger the separation distance the smaller the interaction force. Thus, the above implies the higher frequency peak seen in the upper particle's spectrum represents the resonance frequency of the upper particle while the lower frequency peak represents the driving oscillation due to the interaction force.

According to equation (5), in order to calculate the interaction coefficient $\gamma_{1,2}$, the neutral drag coefficient $\beta_{1,2}$ must be known. This can be accomplished by fitting the FFT spectra as shown in Fig. 3, which yields $\beta_{1,2} = 6$ s$^{-1}$. This is again in reasonable agreement with previously published results (5.0 ± 0.9 for 9.7μm particles (Ref [19]).

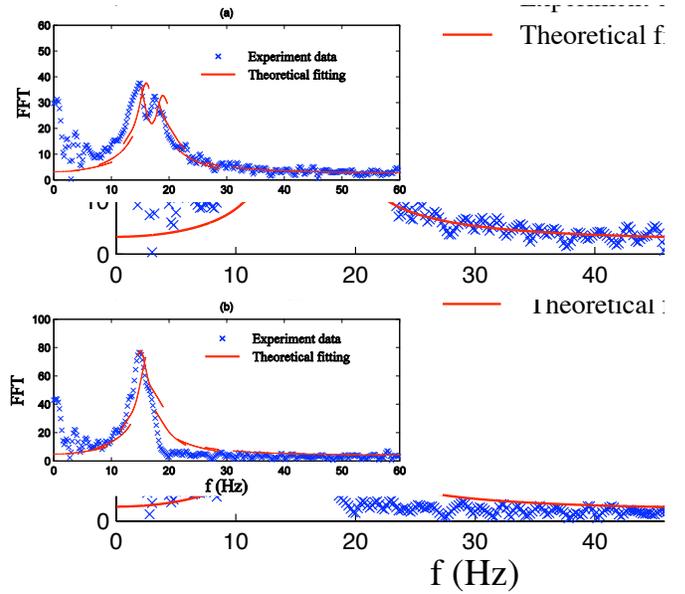

Fig. 3. Experimental FFT data and corresponding theoretical fittings. The data shown in (a) is for the upper particle, while the data shown in (b) is for the lower particle. The neutral drag coefficient is assumed to be $\beta_{1,2} = 6$ s$^{-1}$.

The theoretical fittings shown were derived from the Fourier transformation of the oscillation amplitude, $A = \sum_i a_{0i} e^{-\beta_i t} \cos\omega_{0i} t$, where $a_{0i}$ is the amplitude, $\beta_i$ is the frictional coefficient, and $\omega_{0i}$ is the resonance frequency of the i$^{th}$ peak within the spectrum [14].

After obtaining $\beta_{1,2}$, force constants were then derived using equation (5). These results are shown in Fig. 4. As can be seen, values for $\gamma_1$ center around 3000, except for particle pair p3 where they are approximately 6000. For the same conditions, values of $\gamma_2$ range between 1500 and 2000. (The frequency range was chosen as $\omega_{01,2}^2 - \omega^2 = \omega_{01,2}^2 - \omega_{02,1}^2$, i.e. for $\gamma_1$, $\omega$ is around $\omega_{02}$, while for $\gamma_2$, $\omega$ is around $\omega_{01}$.)

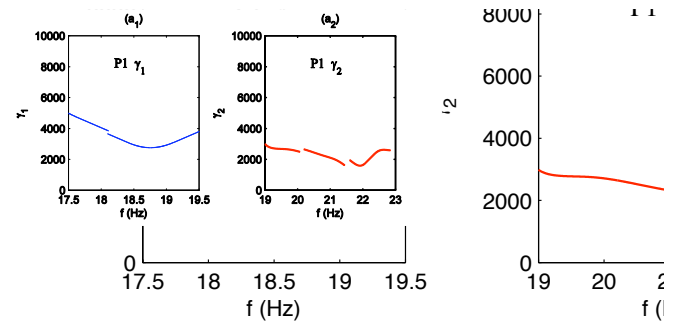

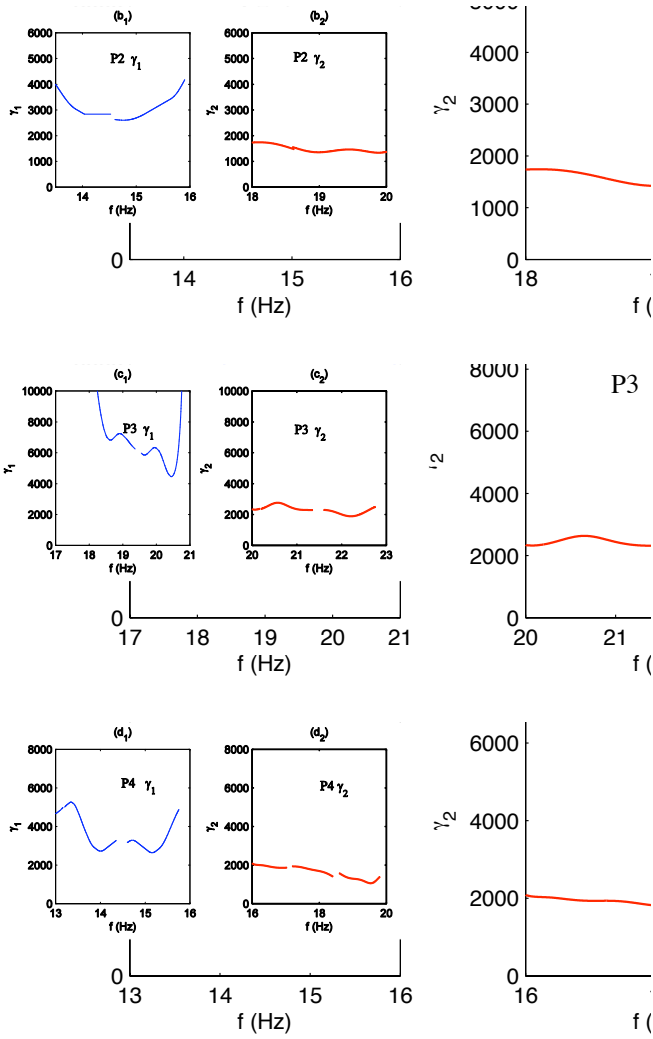

Fig. 4. Interparticle force constants calculated from equation (5) as described in the text. Figure 4($a_1$) − ($a_4$) show the upper particle's force coefficient $\gamma_1$, while Figure 4($b_1$) − ($b_4$) are for the bottom particle's force coefficient $\gamma_2$. p1 − p4 are the particle pairs as defined in the text.

Comparison of the four particle pairs examined show higher $\gamma$ values for smaller particle separation distances representing stronger interaction forces as defined by Eq. 2.

## IV. Discussion

Fig. 4 shows positive values for both $\gamma_1$ and $\gamma_2$, implying the resulting interaction forces are repulsive. Thus, the difference between their values can be related to the anisotropic properties of the sheath.

Unfortunately, before a more comprehensive examination of these anisotropic properties can be developed, the linear force assumption must first be examined in greater detail. As is well known, a linear force approximation, $\gamma_{1,2}|y_2 - y_1|$ can only be assumed valid for small amplitudes. Large vibration amplitudes introduce nonlinear effects (e.g., hysteresis, superharmonics, and subharmonics [15], [16]) creating significant deviations from a linear force approximation. To test the validity of this assumption for the present case, a Yukawa potential was approximated by a quadratic function.

The fitting curve is shown in Fig. 5, where $\mu = |y_2 - y_1| / \lambda_D$, and $\lambda_D$ is the Debye length. In the experiment above, the measured Debye length is $\lambda_D$ = 100 μm [17], the average interparticle separation distance is approximately 240 μm and the highest peak to peak oscillation amplitude is 240 μm. This yields a fitting range of μ = 240 /100, 2Δμ = ± 240 / 100. As can be seen in Fig. 5, calculated residuals indicate that for the current case a linear force approximation is indeed valid.

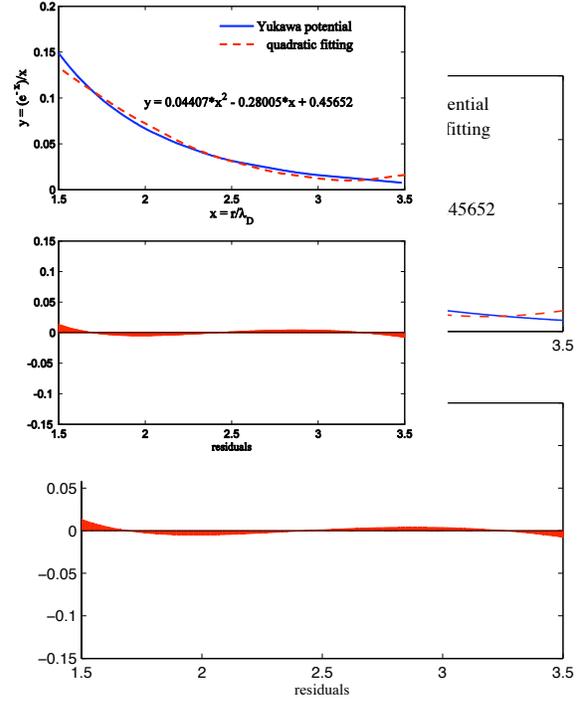

Fig. 5. A quadratic fit to the Yukawa potential. The solid line represents the theoretical Yukawa potential while the dotted line represents a quadratic fit to this potential. Residuals confirm a deviation of around 5% over the range shown.

Additionally, both simulation and experiment [15], [16], [17] have shown the primary conditions necessary for the appearance of nonlinear effects, such as superharmonics, subharmonics, and/or hysteresis, to be low neutral gas pressures and high oscillation amplitudes. Since for the current experiment, both of these were carefully controlled and Fig. 2 shows no evidence of higher harmonics, it is reasonable to assume that Eq. 5 can be considered a valid expression for the current case.

From Eq. 1, the interaction forces exerted on the particles are defined as $F_{12,21} = m_{2,1}\gamma_{2,1}|y_1 - y_2|$. In the current case, the spherical dust particle mass is $m = 4/3\pi r^3 \rho$, where $\rho$ = 1.51 g/cm$^3$, and the dust particle diameter is 8.89 ± 0.09 μm as referenced by the manufacturer. Assuming a worst case scenario, the diameters for the upper and lower particles should therefore fall between (8.89 - 0.09) μm and (8.89 + 0.09) μm. Notice that both $F_{21}$ and $F_{12}$ are positive since $\gamma_{1,2}$ are positive; therefore, each particle exerts a repulsive force on the other. Calculating the ratio $F_{21} / F_{12}$ yields a value greater than 1.5, showing this to be an anisotropic force with the lower particle exerting a much stronger repulsive force on the upper particle than vice versa. This result also explains the strong low frequency peak seen in the FFT spectrum of the upper particle for small particle separation distances, while barely recognizable high frequency peaks appear in the bottom particle's spectrum. As mentioned previously, the source of this anisotropic force is assumed to be the wakefield created by the streaming ions within the sheath. The upper particle is only acted upon by the Yukawa force from the lower particle, while the lower particle sees both the Yukawa force from the upper particle and the wakefield force created by ions passing around the upper particle.

Employing Eq. 5, an estimated experimental error can be derived where the major contribution to this error comes from the ratio R,

$$\frac{|\Delta \gamma|}{\gamma} \geq 2 \frac{|\Delta R|}{R} \approx 2\left(\frac{|\Delta Y_2|}{Y_1} + \frac{|\Delta Y_1|}{Y_2}\right) \approx 4 \frac{|\Delta Y_1|}{Y_1} \qquad (6)$$

For an estimated experimental error of $|\Delta Y| / Y = 5\%$, the error in $\Delta \gamma / \gamma$ will be $\geq \pm 20\%$. The easiest method for reducing this error is to increase the overall signal to noise ratio.

## V. Conclusions

Utilizing a linear force approximation and an attenuated oscillation approach, a new experimental method for probing the interparticle force between two particles within a vertical chain has been developed. Anisotropic effects within the plasma sheath created by the ion streaming effect were directly measured as differences in the forces on the upper and lower particles in the chain. Thus, this experiment shows that, in addition to the previously discovered horizontal confinement effect produced by the ion wakefield, a measurable interaction force in the vertical direction is also present (Ref. 9). It is assumed these forces are also created by the ion wakefield potential. Due to this interaction force, upper particles experience a stronger vertical interaction force than do the lower particles, $F_{21} > 1.5 F_{12}$ as a result of the anisotropic nature of the forces involved. ($F_{21}$ is the interaction force from particle 2, the bottom particle, to particle 1, the upper particle, and vice versa.) It is important to note that the linear force approximation used above is only valid under the condition of small oscillation amplitudes and low neutral gas pressures. Under these conditions and for the experimental parameters discussed, the validity of the linear force approximation was shown to be valid to about ± 5% across the regimes examined.